\documentclass[epj]{webofc}
\usepackage[varg]{txfonts}   
\usepackage{graphicx,color} 
\usepackage{bm}       
\usepackage{amsmath}  
\usepackage{amssymb}
\woctitle{21st International Conference on Few-Body Problems in Physics}
\begin{document}
\title{Reactions with a $^{10}$Be beam to study the one-neutron halo nucleus $^{11}$Be}
\author{K.L.~Jones\inst{1}\fnsep\thanks{\email{kgrzywac@utk.edu}} 
}

\institute{Department of Physics and Astronomy, University of Tennessee, Knoxville, Tennessee 37996, USA.
          }

\abstract{
  Halo nuclei are excellent examples of few-body systems consisting of a core and weakly-bound halo nucleons.  Where there is only one nucleon in the halo, as in $^{11}$Be, the many-body problem can be reduced to a two-body problem. The contribution of the 1s$_{1/2}$ orbital to the ground state configuration in $^{11}$Be, characterized by the spectroscopic factor, $S$, has been extracted from direct reaction data by many groups over the past five decades with discrepant results.  An experiment was performed at the Holifield Radioactive Ion Beam Facility using a $^{10}$Be primary beam at four different energies with the goal of resolving the discrepancy through a consistent analysis of elastic, inelastic, and transfer channels.  Faddeev-type calculations, released after the publication of the experimental results, show that dynamic core excitation in the transfer process can lead to reduced differential cross sections at higher beam energies.  This reduction would lead to the extraction of decreasing values of $S$ with increasing beam energy.  A $^{10}$Be(d,p) measurement at E$_d$ greater than 25~MeV is necessary to investigate the effects of core excitation in the reaction.
}
\maketitle
\section{Introduction}
\label{intro}
Close to particle emission thresholds, nuclei can form clusters reducing an A-body problem into an n-body problem, where A is the mass number representing the sum of the number of neutrons and protons and n is the number of clusters.  An extreme case of this phenomenon can exist close to a single-nucleon emission threshold where A-1 nucleons cluster into a core and the last, weakly-bound nucleon forms a diffuse halo. In this one-nucleon halo, an A-body problem is reduced to a two-body problem, the core, and the halo nucleon.

In addition to proximity to a particle-emission threshold, usually characterized by a small separation energy, a well-formed nuclear-halo system requires small potential barriers.  The last neutron in $^{11}$Be has a separation energy of just, S$_n=0.502$~MeV, compared to a typical value of  $7-8$~MeV for stable isotopes.  The 1/2$^+$ ground state is in large part a result of the lowering of the $n\ell j=2s_{1/2}$ single-particle state.  The lack of a strong Coulomb or centrifugal barrier, leads to a one-neutron halo, as discovered by Tanihata {\it et al.} \cite{Tan88}.

The degree to which the $n\ell j=2s_{1/2}$ single-particle state contributes to the ground-state halo of $^{11}$Be is usually characterized by the spectroscopic factor, $S$.  Spectroscopic factors can be extracted from data taken in direct reaction measurements by comparing differential cross sections with those calculated with a reaction theory.  Commonly, the distorted wave Born approximation (DWBA) is used in the reaction calculation with an $S$ of 1 assumed.  The DWBA calculation is then scaled to the data to extract an experimental value of $S$. The extracted spectroscopic factors are model dependent, with uncertainties entering through the optical model chosen and, especially at low beam energies, the radius and diffuseness of the binding potential.   The adiabatic wave approximation (ADWA) \cite{Joh70} explicitly accounts for deuteron breakup that otherwise could affect extracted spectroscopic factors in a non-trivial way. 

An alternate approach to analyzing direct reaction data is to use a value of $S$ calculated from a structure mode as an input into a reaction model to predict a differential cross section.  Experimental and theoretical cross sections can then be compared directly.  This approach is typically favored when using more sophisticated reaction theory methods that may not scale, such as continuum discretized coupled channels (CDCC) \cite{Aus87} and exact Faddeev-type methods \cite{Fad60, Alt67}.

\section{Spectroscopic factors for $^{11}$Be}
\label{sec-1}
\begin{figure}
\centering
\includegraphics[trim=0.1cm 1cm 0cm 1cm, width=10cm,clip]{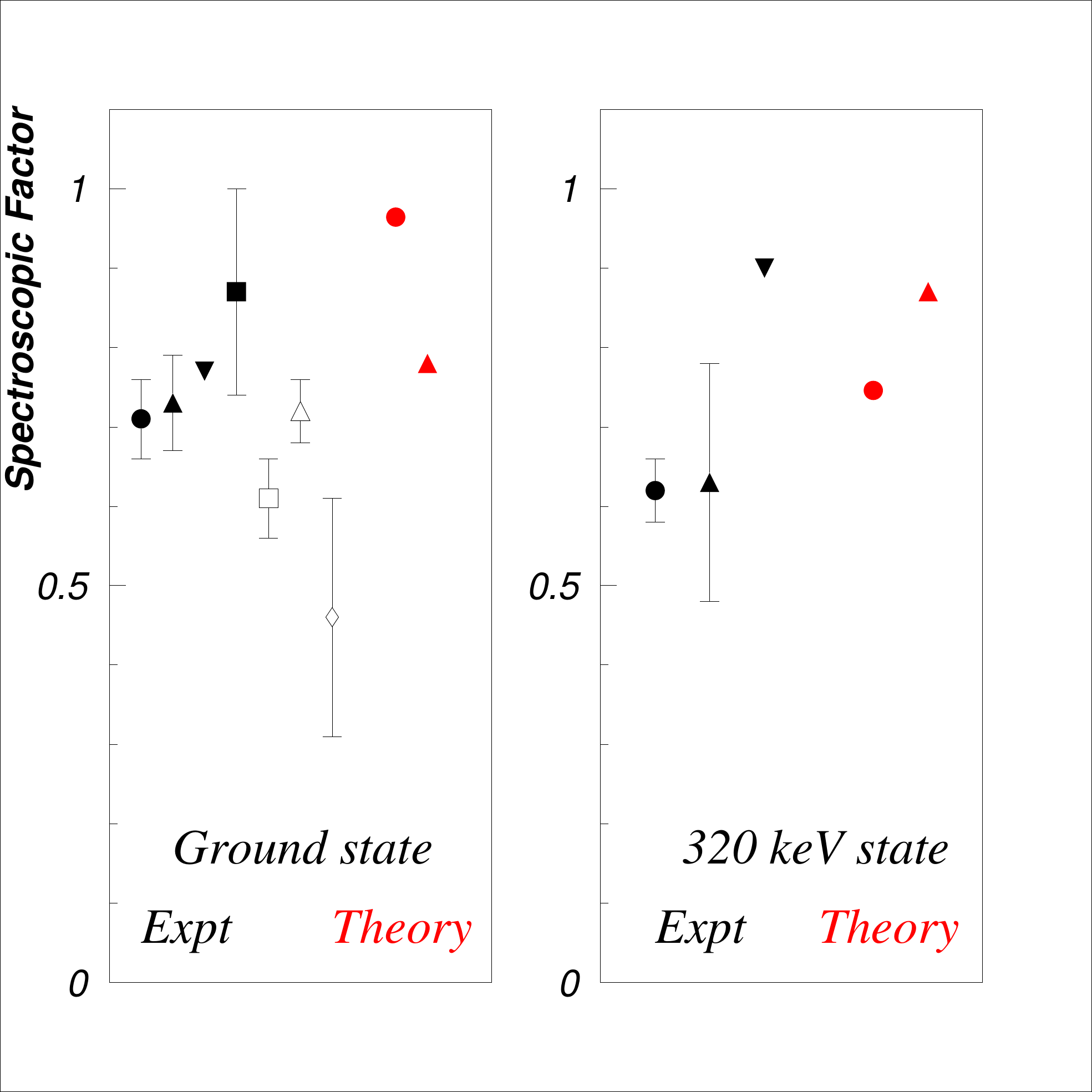}
\caption{Spectroscopic factors extracted from measurement (black) compared to those from theory (red).  The experimental data are from Ref. \cite{Sch13} (circles), Ref. \cite{Aut70} (filled triangles), Ref. \cite{Zwi79} (inverted triangles), Ref. \cite{Aum00} (filled square), Ref. \cite{Pal04} (open square), Ref. \cite{Fuk04} (open triangle), Ref. \cite{Lim07} (diamond).  The theoretical points are from Ref. \cite{Vin95} (red circle) and Ref. \cite{Nun96} (red triangle).  }
\label{fig-1}       
\end{figure}

The structure of $^{11}$Be has been studied using many methods including $\beta$ decay, neutron knockout, and transfer reactions, as summarized by Fortune and Sherr \cite{For12}, as well as many theoretical studies. Figure \ref{fig-1} shows the value of $S$ extracted from direct reaction experiments for the ground and first excited states in $^{11}$Be, compared to calculated values (in red).   For the ground state, the transfer reaction results are all in agreement within error bars.  However, for the first excited state, the value  of $S$ from the measurement of Zwieglinski \cite{Zwi79} (inverted triangles) is significantly larger than the other measurements.  The knockout and Coulomb dissociation measurements give information only about the ground state.  The four measurements shown here gave values of $S$ ranging from $0.73\pm0.13$ \cite{Aum00}(filled square) to $0.46\pm0.15$ \cite{Lim07}(diamond).  
From where do these discrepancies arise?  Were there unknown systematic problems with the experiments, or can the discrepancies be attributed to the different reaction models used in the analysis?  As mentioned before, the choice of both optical model and binding potentials can induce uncertainties in the $S$ extracted from experimental data.  There are other effects that may, or may not, be included in the reaction analysis, such as deuteron break-up, finite range effects and dynamic core excitation.  The former effect can be dealt with by using an ADWA formalism, and finite range reaction codes are generally available.  Core excitation has not generally be included in transfer reaction analyses to date.

The experiment discussed here \cite{Sch12, Sch13} was performed with the goal of providing consistent $^{10}$Be + d data at four different energies, including elastic and inelastic scattering as well as transfer to bound and resonant states.  $S$ was extracted in a consistent fashion, using the same optical potentials and reaction formalisms.  The data were taken at the Holifield Radioactive Ion Beam Facility \cite{Bee11} at Oak Ridge National Laboratory using a primary beam of $^{10}$Be material (which has a half life of over a million years). The hypothesis for these measurements was that the $S$ extracted for each state should be the same, regardless of the beam energy.  Naively, this appears to be a minimum requirement of $S$.

\begin{figure}
\centering
\includegraphics[trim=0.2cm 0.25cm 1.1cm 1.3cm, width=12cm,clip]{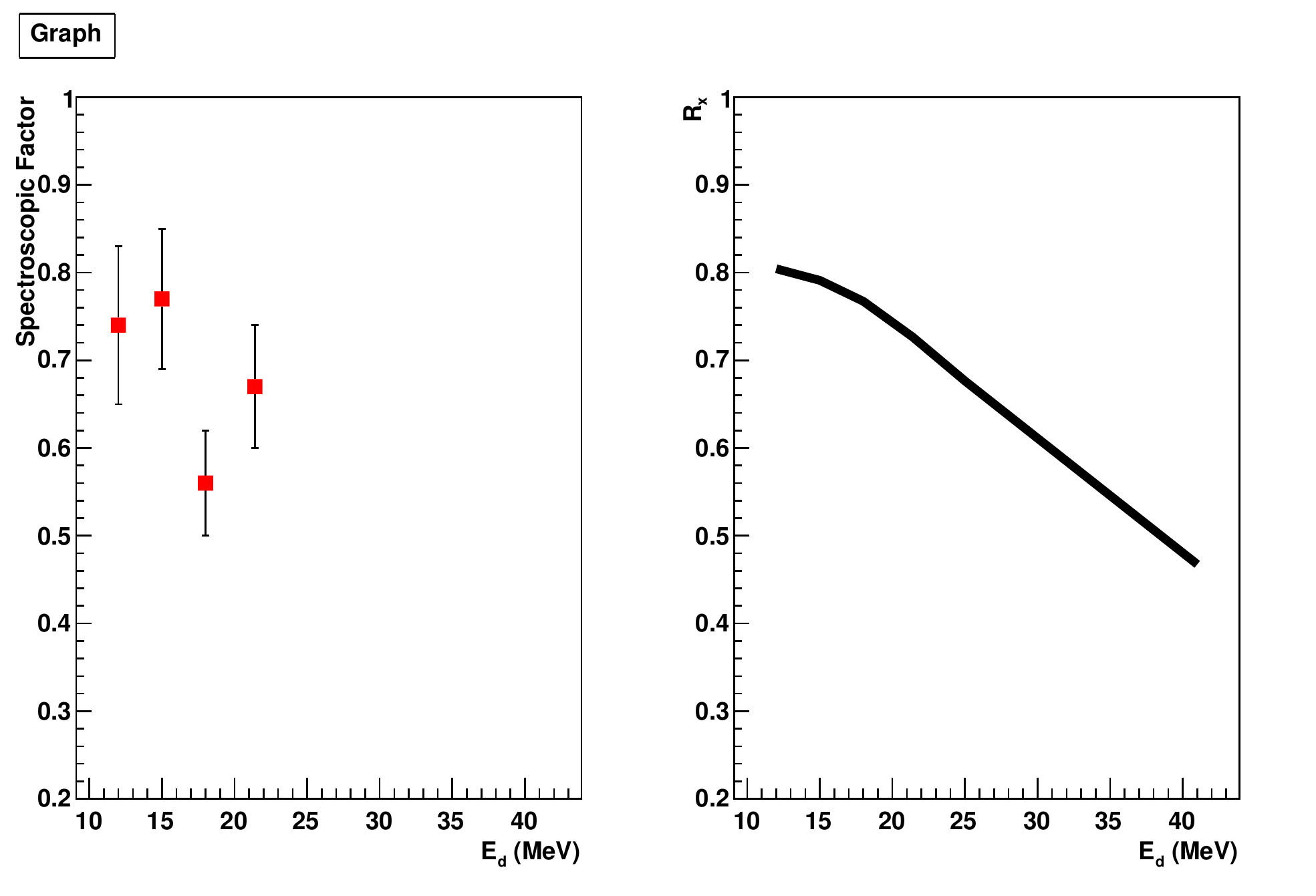}
\caption{Left panel:  Spectroscopic factors for $^{11}$Be$_{gs}$ extracted from the $^{10}$Be(d,p) reaction in inverse kinematics in \cite{Sch13} using the Koning and Delaroche optical potential \cite{Kon03}.  Right panel:  Ratios R$_x$ of differential cross sections for $^{10}$Be(d,p)$^{11}$Be$_{gs}$ calculated including core excitation, adapted from Fig. 5 \cite{Del13}.  }
\label{fig-2}       
\end{figure}

The left panel of Figure \ref{fig-2} shows the values of $S$ extracted for the ground state of $^{11}$Be at E$_{beam}~=~60, 75, 90,$ and 107~MeV (that is equivalent deuteron energy E$_d=12, 15, 18,$ and 21.4~MeV).  These data were analyzed in the finite range ADWA formalism \cite{Joh74, Ngu10}, using the optical potential of Koning and Delaroche \cite{Kon03}.  The bound state parameters used were: radius $r=1.25$~fm, and diffuseness $a=0.65$~fm.  A spin-orbit term with the same geometry as the central interaction and strength V$_{so}$=5.5~MeV was also used.  Other details of the calculations are available in \cite{Sch13}.

The four data points in the left panel of Figure \ref{fig-2} show a consistent value for $S$, within the limitations of the uncertainties.  However, they could equally show a downward trend with increasing beam (or equivalent deuteron) energy.

Faddeev-type calculations of $^{10}$Be + d that included dynamic core excitation were subsequently performed by Deltuva  \cite{Del13} for the four energies used in the experiment .  The core excitation effects he found in that work are shown in the right panel of Figure \ref{fig-2} (the line shown here is the red line in Figure 5 of \cite{Del13}).  R$_x$ is defined as $(d\sigma /d\Omega)_x/(d\sigma /d\Omega)_{SP}$, where $x$ and ${SP}$ are the calculated cross sections with core excitation (in this case in both the n-$^{10}$Be and p-$^{10}$Be channels) and in the single-particle model, respectively.  If the calculation "$x$" exactly represents nature, R$_x$ is the number that would be extracted as $S$ if a single-particle framework, such as DWBA or ADWA, were used to analyze the data.  For neutron-transfer to the ground state of $^{10}$Be Deltuva finds that dynamic core excitation leads to a reduction of cross section, which would be misinterpreted as a smaller value of $S$ in an analysis that depends on scaling DWBA-type reaction calculations to experimental data.  The core excitation effects for transfer to the 1/2$^-$ first excited state are less apparent.

The data shown in Figure \ref{fig-2} are compatible with either a constant $S$, or a reducing value, within error bars over the range of energies used in the experiment.   It is not possible to exclude, nor to confirm, the effects of core excitation within the uncertainties of the current data.  However, the effect is larger at higher beam energies, hence an experiment covering beam energies above E$_d$=25~MeV should be able to distinguish the effects of core excitation.   These effects would be seen in the level of agreement between experimental and theoretical differential cross sections with and without core excitation included.  As the effects on transfer to the first excited state are predicted to be smaller, consistent agreement with transfer to both bound states in $^{11}$Be would be a clear validation of the theoretical description of the process.

\section{Summary}

The one-neutron halo system $^{11}$Be can be considered as a two-body system owing to the low separation energy of the last neutron.  This allows calculations of reactions involving $^{11}$Be to be made using three-body models.  The $^{10}$Be(d,p)$^{11}$Be reaction was performed in inverse kinematics at four different energies.  The data from the elastic, inelastic, and transfer channels were analyzed in a distorted wave and adiabatic framework, as presented in \cite{Sch12, Sch13} and spectroscopic factors were extracted.  The values of $S$ for the ground state of $^{11}$Be are consistent with 0.69$\pm0.06$, for the analysis using the Koning and Deleroche optical potential \cite{Kon03}.

A subsequent theoretical work \cite{Del13} suggested that a reduction in cross section due to core excitation, not taken into account in either DWBA or ADWA analyses, would result in lower values of $S$ being extracted from experimental data at higher beam energies. The current data are not discriminating enough to confirm, or refute, the effects of core excitation in the reaction.  Further measurements of the $^{10}$Be(d,p) reaction at E$_d$ above 25 MeV would be highly desirable and could provide direct evidence of core excitation in the reaction.  Such an experiment would need to be analyzed in a consistent manner to the current data if a comparison were to be made.  The effect on the first excited state in $^{11}$Be is expected to be smaller.  Experimental data covering a range of beam energies and measuring differential cross sections to both bound states would be sensitive to this differential effect of core excitation.

\begin{acknowledgement}
This work was supported by the US Department of Energy, Office of Science, Office of Nuclear Physics under contract number DE-SC001174 and DE-FG02-96ER40983. This research was sponsored in part by the National Nuclear Security Administration under the Stewardship Science Academic Alliances program under contract DE-FG52-08NA28552. The author would like to thank A. Deltuva for providing theoretical data.
\end{acknowledgement}

%
\bibliography{Jones-FB21}
%
%


\end{document}